\documentclass[12pt,aps,showpacs]{revtex4}%
\usepackage{amssymb}
\usepackage{amsmath}
\usepackage{amsfonts}
\usepackage{graphicx}%
\providecommand{\U}[1]{\protect\rule{.1in}{.1in}}
\begin{document}
\title{Gravothermal Catastrophe, an Example}
\author{E.N. Glass }
\affiliation{Physics Department, University of Michigan, Ann Arbor, MI 48109}
\date{29 July 2010}

\begin{abstract}
This work discusses \textit{gravothermal} \textit{catastrophe} in
astrophysical systems and provides an analytic collapse solution which
exhibits many of the \textit{catastrophe} properties. The system collapses
into a trapped surface with outgoing energy radiated to a future boundary, and
provides an example of catastrophic collapse\textit{.}

\end{abstract}

\maketitle

\section{INTRODUCTION}

Since 1968, Lynden-Bell \cite{LBW68,LBE80,LynB98} has illuminated the concept
of \textit{negative} heat capacity in astrophysical systems. He explained
Antonov's theorem \cite{Ant62} which states (roughly) that a spherical
collection of self-gravitating point masses has no global entropy maximum.
Explanation was needed since the most probable state for such a system is at
the maximum of the Boltzmann entropy. Lynden-Bell and Wood considered a
self-gravitating gas sphere. They calculated the energy $E$ and heat capacity
$C_{V}$ of the isothermal sphere. A graph of the sphere's binding energy vs
density contrast showed a local maximum near an inflection point, and the plot
of specific heat vs density contrast showed a stable branch for
\textit{negative} specific heat.

Negative $C_{V}$ can be understood as follows: the virial theorem for
\textit{inverse square} forces of bounded systems relates the average kinetic
energy and average potential energy as
\[
<K.E.>\ =-\frac{1}{2}<P.E.>.
\]
With total energy $E=\ <K.E.>+<P.E.>$ \ one has%
\[
E=-<K.E.>\ \text{negative,}%
\]
but for moving particles $K.E.=\frac{3}{2}Nk_{B}T.$ \ It follows that the
specific heat is negative%
\[
C_{V}=\frac{dE}{dT}=-\frac{3}{2}Nk_{B}.
\]
A negative $C_{V}$ system in contact with a large thermal reservoir will have
fluctuations that add energy and make its transient temperature lower, causing
inward heat flow which will drive it to even lower temperatures. Thus negative
$C_{V}$ systems cannot reach thermal equilibrium.

Lynden-Bell coined the name "gravothermal catastrophe" for stellar systems
undergoing such collapse. To describe the collapse, we quote \cite{LynB68}:
"Conductive transfer of heat from the central region will raise the high
central temperature faster than it raises the lower temperature of the outer
parts. No equilibrium is possible; the center continues to contract and get
hotter, sending out heat to the outer parts."

Astrophysical systems where gravothermal catastrophe may occur \cite{BT03} are
older globular clusters with compact cores, and bright elliptical galaxies
with high central density profiles. Lynden-Bell and Eggleton studied the core
collapse of globular clusters. In particular, they calculated self-similar
collapse features of a gravitating gas sphere. One of the things they learned
from their study was, that to model the formation of a central black hole in a
globular cluster, they needed to go beyond self-similarity and study more
dissipative collapse.

In this work, we present an analytic solution \cite{Gla89} with shear and
radial heat flow (SRH) which models many of the features of catastrophic
collapse. The system collapses into a trapped surface with outgoing energy
radiated to a future boundary.

In the following section, the SRH collapse metric with pressure and density is
developed. The results are given in section III and summarized in section IV.
This is followed by two appendices. (A) contains the original SRH solution
with arbitrary functions, and (B) has mass and trapped surface equations.

\section{COLLAPSING SRH FLUID}

The SRH spacetime \cite{Gla89} is divided into an exterior region covered by
the Vaidya metric and a collapsing interior region with spherically symmetric
metric%
\begin{equation}
g_{\mu\nu}^{\text{SRH}}dx^{\mu}dx^{\nu}=A^{2}dt^{2}-B^{2}dr^{2}-R^{2}%
d\Omega^{2} \label{collapse-met}%
\end{equation}
where $A=A(t,r),$ $B=B(t,r),$ $R=R(t,r),$ and $d\Omega^{2}$ is the metric of
the unit sphere. The stress-energy tensor is given by ($G=c=1$)
\begin{equation}
T^{\mu\nu}=w\hat{u}^{\mu}\hat{u}^{\nu}-p\gamma^{\mu\nu}+q^{\mu}\hat{u}^{\nu
}+\hat{u}^{\mu}q^{\nu}. \label{energy-mom}%
\end{equation}
$p$ is the isotropic pressure, $w$ is the mass-energy density, $\gamma^{\mu
\nu}=g^{\mu\nu}-\hat{u}^{\mu}\hat{u}^{\nu}$, and $q^{\mu}$ is the radial heat
flow vector orthogonal to $\hat{u}^{\mu}$. Time is comoving with $\hat{u}%
^{\mu}\partial_{\mu}=A^{-1}\partial_{t}$, $\hat{u}^{\mu}\hat{u}_{\mu}=1$.\ We
use notation of Taub \cite{Tau69} for the mass-energy density $w$. Taub's
purpose was to distinguish mass-energy density from proper mass-density $\rho
$, where $w=\rho(1+\epsilon)$ and $\epsilon$ is specific internal energy.

The original SRH solution, given in Appendix A, provides the following metric
functions and physical scalars (overdots denote $\partial/\partial t$ and
primes denote $\partial/\partial r$).
\begin{subequations}
\begin{align}
R(t,r)  &  =e^{t/t_{0}}[(\beta_{0}^{2}/2)e^{-4t/t_{0}}+(r+r_{0})^{2}%
]^{1/2},\label{R-eq}\\
A(t,r)  &  =\alpha_{0}\dot{R},\\
B(t,r)  &  =\beta_{0}/R.
\end{align}
Metric functions $A$ and $B$ should be dimensionless and $R$ should have
dimensions of length. We write $A=\alpha_{0}(\dot{R}/c)$ so it is clear that
$\alpha_{0}$ is dimensionless. The other constants have the following
dimensions: $\dim(t_{0})=time,$ \ $\dim(\beta_{0})=length$, \ $\dim
(r_{0})=length$.

The heat flow vector and scalar are given by $4\pi q_{\mu}dx^{\mu}=Qdr$. The
heat flow scalar is
\end{subequations}
\begin{equation}
Q=-\frac{1}{\alpha_{0}}(\frac{r+r_{0}}{R^{3}})e^{2t/t_{0}}. \label{Q-eqn}%
\end{equation}
The pressure is
\begin{equation}
8\pi p=\frac{(r+r_{0})^{2}e^{4t/t_{0}}}{\beta_{0}^{2}R^{2}}\left[  \frac
{R^{2}+\beta_{0}^{2}e^{-2t/t_{0}}}{R^{2}-\beta_{0}^{2}e^{-2t/t_{0}}}\right]
-\frac{(1+1/\alpha_{0}^{2})}{R^{2}},
\end{equation}
and the mass-energy density is
\begin{equation}
8\pi w=\frac{1}{R^{2}}\left[  1-\frac{1}{\alpha_{0}^{2}}-\frac{(r+r_{0})^{2}%
}{\beta_{0}^{2}}e^{4t/t_{0}}\right]  -\frac{2}{\beta_{0}^{2}}e^{2t/t_{0}}.
\end{equation}
$w$ is negative when%
\begin{align*}
\left[  \alpha_{0}^{2}-1-\frac{\alpha_{0}^{2}(r+r_{0})^{2}}{\beta_{0}^{2}%
}e^{4t/t_{0}}\right]  -\frac{2\alpha_{0}^{2}R^{2}}{\beta_{0}^{2}}e^{2t/t_{0}}
&  <0\\
\alpha_{0}^{2}\beta_{0}^{2}-\beta_{0}^{2}-\alpha_{0}^{2}(r+r_{0}%
)^{2}e^{4t/t_{0}}-2\alpha_{0}^{2}R^{2}e^{2t/t_{0}}  &  <0\\
\text{or \ \ \ \ \ \ \ \ \ \ \ \ \ \ \ \ \ }\beta_{0}^{2}+3\alpha_{0}%
^{2}(r+r_{0})^{2}e^{4t/t_{0}}  &  >0.
\end{align*}
This inequality always holds, and so $w$ is negative.

\section{COLLAPSE RESULTS}

\subsection{End stage of collapse}

For small $\beta_{0}$, $R$ goes as%
\begin{equation}
R\simeq e^{t/t_{0}}(r+r_{0}).
\end{equation}
The pressure and mass-energy density, at late times and for small $\beta_{0}
$, go as
\begin{align}
8\pi p  &  \simeq\frac{e^{2t/t_{0}}}{\beta_{0}^{2}},\label{late-p}\\
8\pi w  &  \simeq-3\frac{e^{2t/t_{0}}}{\beta_{0}^{2}}. \label{late-w}%
\end{align}
At late times, the magnitude of the mass-energy density increases
exponentially, and the fluid approaches a photon gas with equation of state
\begin{equation}
p=\frac{1}{3}\mid w\mid.
\end{equation}

Negative mass-energy density $w$ is linked to the \textit{gravothermal
catastrophe}. The collapsing fluid has a negative specific heat which derives
from negative internal energy $\epsilon$. While both $w$ and $\epsilon$ are
negative, the proper mass density, $\rho=w/(1+\epsilon)$, remains positive. To
separately compute $\epsilon$ would require a complete thermodynamic analysis.
One would need a causal relativistic description of thermodynamics such as the
Israel-Stewart `second order' type theory \cite{IS79,Cal98}. This has been
left for future work.

As a model of gravitational dyamics, Chavanis et al \cite{CRS02} studied the
collapse of a gas of self-gravitating Brownian particles in a closed sphere.
For catastrophic collapse in the microcanonical ensemble, they found the mass
density to go as%
\begin{equation}
\rho\simeq\rho_{0}(r_{0}/r)^{2.21} \label{brown-part-dens}%
\end{equation}
and further, in a stability study, they found the perturbation $\delta
\rho/\rho$ has a "core-halo" structure (hinting at central black hole
formation). Without developing the thermodynamics of a collapsing SRH fluid,
the density approximation in Eq.(\ref{brown-part-dens}) can be used to compute
internal energy from $\epsilon=w/\rho-1$. With $w$ at late times given in
Eq.(\ref{late-w}), we find%
\begin{equation}
\epsilon\simeq-const\text{ }\frac{e^{2t/t_{0}}}{\rho_{0}\beta_{0}^{2}}%
(r/r_{0})^{2.21}-1
\end{equation}
When the density is given a temperature profile (in the canonical ensemble
$\rho\sim T^{-1/2}$), then $\epsilon=\epsilon(T)$ yields negative specific
heat $c_{V}=d\epsilon/dT$.

\subsection{Trapped Surface}

From Eq.(\ref{m-curv}) the Misner-Sharp mass within a $t=const$, $r=const$
2-surface is%
\begin{equation}
2m=R\left[  1+\frac{1}{\alpha_{0}^{2}}-\frac{(r+r_{0})^{2}}{\beta_{0}^{2}%
}e^{4t/t_{0}}\right]  .\label{mass}%
\end{equation}
Null rays entering and leaving the 2-surface are described by the expansions
of their respective generators. Equations (\ref{div-n}) and (\ref{div-l}) show
that both null generators have non-negative expansions and so a trapped
surface exisits at $R=2m$. At the trapped surface, Eq.(\ref{mass}) sets a
value for constants $\alpha_{0}$ and $\beta_{0}$.%
\begin{equation}
\frac{\beta_{0}^{2}}{\alpha_{0}^{2}}=(r_{\text{trap}}+r_{0})^{2}%
e^{4t_{\text{trap}}/t_{0}}.
\end{equation}
The expression for $R$, Eq.(\ref{R-eq}), yields%
\begin{equation}
R_{\text{trap}}^{2}=(r_{\text{trap}}+r_{0})\beta_{0}\frac{(2+\alpha_{0}^{2}%
)}{2\alpha_{0}}.
\end{equation}

We see from Eq.(\ref{Q-eqn}) that, at late times, the heat flow scalar goes to
zero%
\begin{equation}
Q\simeq-\frac{1}{\alpha_{0}}\frac{1}{(r+r_{0})^{2}e^{t/t_{0}}}\rightarrow0.
\end{equation}
The heat flow shuts off as the fluid collapses into the trapped surface.

\subsection{Rate of collapse}

The rate of collapse scalar is $\Theta=\nabla_{\mu}\hat{u}^{\mu}%
=-1/(\alpha_{0}R)$. We again quote Lynden-Bell: "During the gravothermal
catastrophe ... the center continues to constrict and get hotter, giving out
heat to the outer parts, but the temperature difference increases and drives
the collapse onwards still faster." At distances just beyond the trapped
surface%
\begin{equation}
\Theta\simeq-\frac{e^{t/t_{0}}}{\alpha_{0}\beta_{0}}.
\end{equation}
At late times the rate of collapse increases exponentially. This is a
necessary component for a model of catastrophic collapse.

\section{SUMMARY}

An analytic solution of Einstein's equations for dissipative collapse has been
presented. The original SRH solution contains arbitrary functions of time
which have been chosen here to provide an explicit solution. The system
collapses into a trapped surface with outgoing energy radiated to a future
asymptotic boundary. The collapsing fluid has negative mass-energy, which has
been related to negative specific heat. Lynden-Bell has linked collapse with
negative heat capacity to gravothermal catastrophe. The SRH collapse has many
features that model the gravothermal catastrophe.

\appendix{}

\section{Original SRH Solution}

This solution is given in equations (14), (15), and (16) of \cite{Gla89}
[there is an error in the first term of Eq.(16). $\beta_{0}^{2}h_{1}$ should
be ($\beta_{0}^{2}/2)h_{1}$]. The metric components are, with arbitrary
functions $h_{1}(t)$ and $h_{2}(t)$
\begin{subequations}
\label{one}%
\begin{align}
A(t,r)  &  =\alpha_{0}\dot{R},\\
B(t,r)  &  =\beta_{0}/R,\\
R(t,r)  &  =[(\beta_{0}^{2}/2)h_{1}+h_{1}^{-1}(r+h_{2})^{2}]^{1/2}.
\end{align}
The match of $g^{\text{SRH}}$ to exterior Vaidya was done in \cite{Gla89}.

The pressure, with parameters and arbitrary functions unchosen, is
\end{subequations}
\begin{align}
8\pi p  &  =\frac{R^{2}}{\beta_{0}^{2}}\left[  (\frac{R^{\prime}}{R}%
)^{2}+2\frac{\dot{R}^{\prime}}{\dot{R}}\frac{R^{\prime}}{R}\right]  -\frac
{1}{\alpha_{0}^{2}\dot{R}^{2}}\left[  (\frac{\dot{R}}{R})^{2}\right]
-\frac{1}{R^{2}}\nonumber\\
&  =\frac{1}{\beta_{0}^{2}}\left[  (R^{\prime})^{2}+(R^{2})^{\prime}\frac
{\dot{R}^{\prime}}{\dot{R}}\right]  -\frac{1}{R^{2}}(1+\frac{1}{\alpha_{0}%
^{2}})\nonumber\\
&  =-\frac{(\frac{r+h_{2}}{h_{1}})^{2}}{\beta_{0}^{2}R^{2}}+\frac
{2(\frac{r+h_{2}}{h_{1}})[(\frac{r+h_{2}}{h_{1}})-\frac{\dot{h}_{2}}{\dot
{h}_{1}}]}{\beta_{0}^{2}[R^{2}-\beta_{0}^{2}h_{1}-2(\dot{h}_{2}/\dot{h}%
_{1})(r+h_{2})]}-\frac{(1+1/\alpha_{0}^{2})}{R^{2}}%
\end{align}
with mass-energy density
\begin{align}
8\pi w  &  =\frac{1}{\alpha_{0}^{2}\dot{R}^{2}}\left[  -(\frac{\dot{R}}%
{R})^{2}\right]  -\frac{R^{2}}{\beta_{0}^{2}}\left[  2\frac{R^{\prime\prime}%
}{R}+3(\frac{R^{\prime}}{R})^{2}\right]  +\frac{1}{R^{2}}\nonumber\\
&  =\frac{1}{\alpha_{0}^{2}R^{2}}\left[  \alpha_{0}^{2}-1-\frac{\alpha_{0}%
^{2}(r+h_{2})^{2}}{\beta_{0}^{2}h_{1}^{2}}\right]  -\frac{2}{\beta_{0}%
^{2}h_{1}}.
\end{align}
The heat flow vector and scalar are given by $4\pi q_{\mu}dx^{\mu}=Qdr$,
\begin{align}
Q  &  =-\frac{1}{\alpha_{0}}\frac{R^{\prime}}{R^{2}}\\
&  =-\frac{1}{\alpha_{0}R^{2}}\left[  \frac{r+h_{2}}{h_{1}}\right]  .\nonumber
\end{align}
For the particular case above, we choose
\begin{equation}
h_{1}=e^{-2t/t_{0}},\text{ \ }h_{2}=r_{0}.
\end{equation}

\section{MASS and TRAPPED SURFACE}

The collapse metric is spanned by the tetrad
\begin{align*}
\hat{u}_{\mu}dx^{\mu}  &  =Adt,\ \ \hat{r}_{\mu}dx^{\mu}=Bdr,\\
\hat{\vartheta}_{\mu}dx^{\mu}  &  =Rd\vartheta,\ \ \hat{\varphi}_{\mu}dx^{\mu
}=R\text{sin}\vartheta d\varphi
\end{align*}
such that%
\begin{equation}
g_{\mu\nu}=\hat{u}_{\mu}\hat{u}_{\nu}-\hat{r}_{\mu}\hat{r}_{\nu}%
-\hat{\vartheta}_{\mu}\hat{\vartheta}_{\nu}-\hat{\varphi}_{\mu}\hat{\varphi
}_{\nu}.
\end{equation}

The unique mass \cite{MS64} $m$ within $t=const$, $r=const$ 2-surfaces is
\begin{equation}
2m=R^{3}R_{\alpha\beta\mu\nu}\hat{\vartheta}^{\alpha}\hat{\varphi}^{\beta}%
\hat{\vartheta}^{\mu}\hat{\varphi}^{\nu}=R[1+\dot{R}^{2}/A^{2}-(R^{\prime
})^{2}/B^{2}].\label{m-curv}%
\end{equation}
The metric is Petrov type \textbf{D} and the only non-zero Weyl tensor
component $\Psi_{2}$ is invariantly expressed as $48(\Psi_{2})^{2}%
=C_{\alpha\beta\mu\nu}C^{\alpha\beta\mu\nu}$. The two principal null vectors
of metric (\ref{collapse-met}), normal to ($\vartheta,\varphi$) 2-surfaces,
are%
\begin{align}
l^{\mu}\partial_{\mu} &  =A^{-1}\partial_{t}+B^{-1}\partial_{r}\label{L-op}\\
n^{\mu}\partial_{\mu} &  =A^{-1}\partial_{t}-B^{-1}\partial_{r}\label{N-op}%
\end{align}
with respective expansions%
\begin{equation}
l_{;\mu}^{\mu}=\frac{\sqrt{2}}{R}(\frac{\dot{R}}{A}+\frac{R^{\prime}}%
{B}),\text{ \ \ }n_{;\mu}^{\mu}=\frac{\sqrt{2}}{R}(\frac{\dot{R}}{A}%
-\frac{R^{\prime}}{B}).\label{divs}%
\end{equation}
Consider a 2-surface $S$ generated by $l^{\mu}$ and $n^{\mu}$. If both null
generators converge, i.e. expansions $l_{;\mu}^{\mu}$ and $n_{;\mu}^{\mu}$
have the same positive sign, then the 2-surface is trapped \cite{HE73}. When
$R=2m$, Eq.(\ref{m-curv}) provides
\[
\frac{\dot{R}}{A}=\frac{R^{\prime}}{B}%
\]
which implies
\begin{equation}
n_{;\mu}^{\mu}=0,\text{ \ }l_{;\mu}^{\mu}=\frac{2\sqrt{2}\dot{R}}%
{RA}.\label{div-n}%
\end{equation}
Using $A=\alpha_{0}\dot{R}$, the expansion of $l^{\mu}$ is%
\begin{equation}
l_{;\mu}^{\mu}=\frac{2\sqrt{2}}{\alpha_{0}R}.\label{div-l}%
\end{equation}
When $\alpha_{0}>0$ both expansions are non-negative signifying a trapped
surface at $R=2m$.

\end{document}